\relax
\documentclass{article} 
\usepackage{ijcai18}  
\usepackage[table]{xcolor}
\usepackage{bm}
\usepackage{comment}
\usepackage{times}  
\usepackage{helvet}  
\usepackage{courier}  
\usepackage{url}  
\usepackage{graphicx}  
\usepackage{booktabs}       
\usepackage{amsfonts}       
\usepackage{nicefrac}       
\usepackage{microtype}      
\usepackage{amsmath}
\usepackage{amssymb}
\usepackage{subfigure}
\usepackage{algorithm}
\usepackage{algorithmic}
\usepackage{accents}
 \usepackage{multirow}
  \usepackage{multicol}
 \usepackage{booktabs}
\usepackage[small]{caption}
\usepackage{booktabs}

\makeatletter
\def\wideubar{\underaccent{{\cc@style\underline{\mskip10mu}}}}
\def\Wideubar{\underaccent{{\cc@style\underline{\mskip8mu}}}}
\makeatother

\makeatletter
\def\widebar{\accentset{{\cc@style\underline{\mskip10mu}}}}
\def\Widebar{\accentset{{\cc@style\underline{\mskip8mu}}}}
\makeatother

\begin{document}
\title{Weakly Supervised Audio Source Separation\\via Spectrum Energy Preserved Wasserstein Learning}
\author{
Ning Zhang$^{1}$,
Junchi Yan\thanks{Junchi Yan is the correspondence author. The work is partially supported by NSFC 61602176, U1609220, 61672231. Part of the work was done when the first author is with IBM Research -- China.}$^2$,
Yuchen Zhou$^1$\\
$^1$ IBM Research -- China, Beijing, P.R. China\\
$^2$ Shanghai Jiao Tong University, Shanghai, P.R. China\\
%
ningzh6610@gmail.com, yanjunchi@sjtu.edu.cn, zhouyuc@cn.ibm.com
}
\maketitle
\begin{abstract}
Separating audio mixtures into individual instrument tracks has been a standing challenge. We introduce a novel weakly supervised audio source separation approach based on deep adversarial learning. Specifically, our loss function adopts the Wasserstein distance which directly measures the distribution distance between the separated sources and the real sources for each individual source. Moreover, a global regularization term is added to fulfill the spectrum energy preservation property regardless separation. Unlike state-of-the-art weakly supervised models which often involve deliberately devised constraints or careful model selection, our approach need little prior model specification on the data, and can be straightforwardly learned in an end-to-end fashion. We show that the proposed method performs competitively on public benchmark against state-of-the-art weakly supervised methods.
\end{abstract}
\section{Introduction}
\subsection{Background}
Audio source separation (ASS) of mixed sources has long been a challenging problem, from which music source separation is a domain where considerable attentions have been attracted. Music source separation has demonstrated its tremendous potential values in various applications, such as music upmixing, audio restoration, music edit, music information retrieval, among others \cite{Liutkus2015Scalable}. In fact, music source separation is challenging because it is an ill-posed inverse problem. Typically we have more signals to estimate than the number of signals we observed. As most music tracks are mono or stereo recordings, the task is to recover the sources of all its constituent instruments. This will be an underdetermined problem if no further regularization scheme is enforced.

Researchers have achieved successive breakthroughs on the music source separation techniques suited for specific occasions. Among these techniques the guided source separation, which means algorithms employ some information about the audio sources or the acoustic mixing process, has shown promising performance for real-world practices. A few of these separation algorithms are conducted in spatial domain~\cite{Cardoso1998Multidimensional}, but most methods switched to the time-frequency domain by means of complex-valued short time Fourier transform (STFT)~\cite{Vincent2014From} or constant-Q transform (CQT)~\cite{Fitzgerald2011Shifted}.

\subsection{Related Work}
Existing music source separation methods can be in general categorized into supervised approaches and unsupervised ones. For the supervised setting, the separated sources are assumed available as labeled data together with the mixture data. This direction has been recently dominated by neural network based methods. Due to the powerful modeling capability of the deep neural networks, these models achieve improvement in most audio source separation fields. \cite{uhlich2015deep} build a deep neural network (DNN) which takes magnitude spectrograms of mixture signal as input and tried to extract individual instruments from music. \cite{nugraha2016multichannel} propose to combine DNN with the classical Gaussian model to perform multichannel audio source separation. They further explore different distance metrics for the training of DNNs, including the Itakura-Saito divergence, Kullback-Leibler, Cauchy, mean squared error, and phase-sensitive cost functions. \cite{chandna2017monoaural} demonstrate the superiority of convolutional neural network (CNN) in monoaural audio source separation. \cite{huang2015joint} adopt a deep recurrent neural networks to separate monaural sources. \cite{uhlich2017improving} employ a data augmentation technique during training and blend an ensemble four DNNs and LSTM (long-short term memory) networks to do music audio source separation. Their blending model ranked the first place in the MUS task of the 2016 Signal Separation Evaluation Campaign~\cite{liutkus20172016}.

The performance of these models largely depend on the assumption that there are large amount of labelled data for training, which is unrealistic in many applications. This restricts their applicability to real-world problems.

The unsupervised methods are mostly trying to solve the ill-posed inverse problem under different regularizations. The Independent Component Analysis (ICA) \cite{Cardoso1998Multidimensional} or Sparse Component Analysis (SCA)~\cite{Comon2010Handbook} based methods assume the source STFT coefficients have a stationary non-gaussian distribution~\cite{Vincent2007Complex}. The local Gaussian modelling (LGM) based methods~\cite{Liutkus2011Gaussian} assumed that the vectors of STFT coefficients of the source spatial images have a zero-mean nonstationary Gaussian distribution. Compared to totally unsupervised separation, combining different level of weakly supervised information about the sources leads to improved performance in practice.  The non-negative matrix factorization (NMF) based methods~\cite{Ozerov2010Multichannel} employ the global structure information of sources and demonstrated impressive performance. The REPET algorithm \cite{rafii2013repeating} takes advantage of the prior that most musical background is repetitive whereas the vocal signal is not.

These unsupervised and weakly supervised methods are built on specific and adhoc assumptions, which limit their generalization capability. Besides, these methods require expensive computation and high processing time, making them difficult to use for real-world applications. Last but not least, state-of-the-art unsupervised models, which in essence are weakly supervised models, e.g. kernel additive model (KAM) \cite{liutkus2014kernel} and OZE \cite{ozerov2012general} all heavily rely on a precise and correct prior knowledge on the separated sources, the deliberate constraint design and kernel selection prevent them from user-friendly. More importantly, in many real-world applications, such prior knowledge is either hard to collect or difficult to be converted to a mathematical formula.
\begin{figure}[tb!]
\begin{center}
\includegraphics[width=0.9\linewidth]{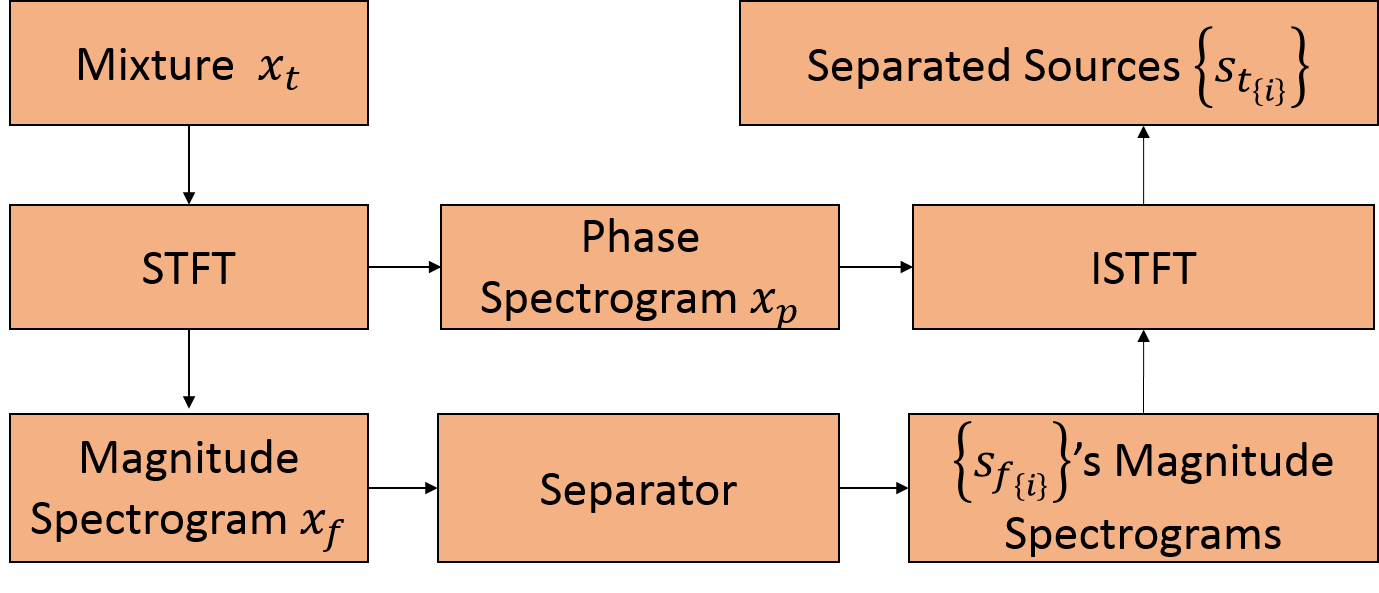}
\end{center}
\caption{Working flow of music audio source separation.}
\label{fig:flow}
\end{figure}
\subsection{Approach Overview}
This paper aims to explore weakly supervised methods as seeing their potential for real-world applications. We pursue a more general approach aiming at relying less on constraints, assumptions, prior knowledge on the source data as done by previous methods \cite{liutkus2014kernel,ozerov2012general}. Specifically our model is based on two mild assumptions making them inherently more general: i) the signal's spectrum energy is preserved after separation which has been a well accepted concept in literature~\cite{sandsten2016time}; ii) the separated sources are in the same distribution with other source data, e.g. separated drum sources from music shall lie in its own distribution which can be approximated by a few (other) real-world drum samples -- which are more easier to collect than in the supervised case as we require no one-to-one correspondence between the separated source and its mixture. In fact, the above two constraints can be complementary to each other as the first one enforces global consistency while the latter pays more attention to its own distribution with no interlock to other separated sources in learning.

From the optimization perspective, the second assumption above can be mathematically converted to the Wasserstein distance which has been proven robust metric to measure the distance between two distributions~\cite{arjovsky2017wasserstein}. In contrast to the KL-divergence that requires strict match between two probability distributions and does not consider how close two samples are but only their relatively probability, which is sensitive to sample noise and outliers, Wasserstein distance is more sensitive to the underlying geometry structure of samples and more robust to noise/outliers. In addition, the first assumption can be explicitly accounted by an analytical energy preservation term in the loss function. To make the Wasserstein distance more tractable, in this paper we turn to the generative adversarial nets (GANs)~\cite{Goodfellow2014Generative} technique. More specifically, the Wasserstein distance minimization problem can be solved by a minmax gaming procedure if we treat our separation model as the generator and introduce an additional discriminator that tries to distinguish the fake generated sources from the real ones. As a result, the efforts boil down to devising an effective generator (see Figure \ref{fig:generator}) and discriminator (see the discriminator modules in Figure \ref{fig:overview}). For the discriminator structure, skip connection is used in this paper to improve its capacity. More importantly, we employ $n$ independent discriminators to account for their corresponding sources respectively. The global consistency is synergically complemented by the energy preservation term. In very recent preprint works, there are also GAN based source separation networks. The SVSGAN in \cite{SVSGAN17} turns to a supervised approach whereby GAN technique is additionally used as an enhancement. Another work \cite{SubakanArxiv17} turns to unconditional GAN for unsupervised learning of different separated tracks, for the purpose of directly learning their different distributions using Wasserstein distance. In contrast, our model is based on the conditional GAN which can directly handle input source for separation, and our end-to-end learning algorithm is totally different from their piece-by-piece ad-hoc method.

To better formulate the problem, we assume the number of sources $n$ for separation is fixed and known, and we further assume all the training and testing data are associated with a same $n$ for separation. This assumption is realistic for real-world applications and has been widely adopted by existing works on music source separation~\cite{Vincent2014From}.

\subsection{Contribution}
In a nutshell, the highlights of this paper are:

1) We explore a new paradigm (a similar concurrent idea appears in \cite{StollerICASSP18}) for weakly supervised source separation, which involves a deep network learned by a general loss function which rely less on prior knowledge on each of the source data for separation, compared with existing weakly supervised models~ \cite{liutkus2014kernel,ozerov2012general}. The direct benefit is for its end-to-end learning capability.

2) We present a novel and general loss involving i) the Wasserstein distance between the distributions of the separated source samples and the real-world source samples; ii) the spectrum energy preservation constraint which can be a complementary to each source's Wasserstein loss, as it imposes global interlock and consistency among the separated sources compared with the mixed signal. We explore the joint use of these two complementary losses to address the highly under-determined problem.

3) Our learning paradigm and loss function, lead to competitive performance on music separation benchmark compared with state-of-the-art weakly supervised methods. In contrast to \cite{liutkus2014kernel} needing careful kernel model selection, and \cite{ozerov2012general} calling for deliberate constraint design, our model is easy to implement and more general.
\section{Weakly Supervised Wasserstein Learning for Audio Source Separation (ASS)}
\subsection{Preliminaries on ASS} We explain the energy preserved Wasserstein learning for separating the mixture audio $x_{t}$ into sources $\{s_{t_{i}}\}_{i=1...n}$. Here $n$ denotes the number of sources. The overall data flow is shown in Figure~\ref{fig:flow}. Firstly, the mixture audio $x_{t}$ is segmented into overlapped segments of time context $C$, on which the short time Fourier transform (STFT) is computed. The resulting magnitude spectrograms $x_{f}(\omega)$ are passed through the separator, which outputs the estimate $\{\hat{s}_{f_{i}}(\omega)\}_{i=1...n}$ for each of the separated sources $i$. Here $\omega$ is the time indices. These estimates, along with the computed phase of the mixture, are transformed through an inverse STFT to recover the audio signals $\hat{x}_{t}$ corresponding to the separated sources.
\subsection{Energy Preserved Wasserstein Learning}
As mentioned in the introduction, the loss function involves i) the energy preservation term to restrict the separated sources's total energy is close to the mixed one; ii) the distribution distance term hoping the separated sources are similar to other real-world sources in the same category.
\subsubsection{Spectrum energy preservation}
This constraint is widely accepted and used in signal processing literature e.g.~\cite{sandsten2016time}, which can be written by:
\begin{equation}\notag
\sum_{i=1}^{n} \|\hat{s}_{f_{i}}(\omega)\|_{2}^{2} = \|x_{f}(\omega)\|_{2}^{2},
\end{equation}
\subsubsection{Wasserstein distance between distributions}
We describe the Wasserstein distance in the language of GANs. It involves two losses $L_g$, $L_d$ for generator and discriminator respectively, as written by \cite{Goodfellow2014Generative}:
\begin{small}
\begin{equation}\label{eq:gans}
\left\{ \begin{array}{ll}
L_{d}=&-\mathbb{E}_{\mathcal{S}_{f} \sim \mathbb{P}(\mathcal{S}_{f})}(D_{\theta}(\mathcal{S}_{f}))+\mathbb{E}_{x_{f}(\omega) \sim \mathbb{P}(x_{f})}(D_{\theta}(G_w(x_{f}(\omega)))) \\
&+ \lambda \mathbb{E}_{\tilde{\mathcal{S}}_{f}(\omega)\sim \mathbb{P}(\tilde{\mathcal{S}}_{f}(\omega))}\left[\left(\|\nabla_{\tilde{\mathcal{S}}(\omega)}D_{\theta}(\tilde{\mathcal{S}}_{f}(\omega))\|_{2}-1\right)^2\right]\\
\\
L_{g}=&-\mathbb{E}_{x_{f}(\omega) \sim \mathbb{P}(x_{f})}\left(D\left(G_w\left(x_{f}(\omega)\right)\right)\right)\\
&+ \lambda \mathbb{E}_{\tilde{\mathcal{S}}_{f}(\omega)\sim \mathbb{P}(\tilde{\mathcal{S}}_{f}(\omega))}\left[\left(\|\nabla_{\tilde{\mathcal{S}}_{f}(\omega)}D_{\theta}(\tilde{\mathcal{S}}_{f}(\omega))\|_{2}-1\right)^2\right]\\
 \end{array}
 \right.
\end{equation}
\end{small}
where $\theta$ and $w$ are the model parameter for the generator and discriminator, respectively. While $\mathcal{S}_{f} = \{s_{f_{1}}, ... , s_{f_{n}}\}$ denotes $n$ sources, and $\lambda$ is gradient penalty weight. $\tilde{\mathcal{S}}(\omega) \sim \mathbb{P}(\tilde{\mathcal{S}}(\omega))$ are points sampled uniformly along straight lines between pairs of points sampled from the data distribution $\mathbb{P}(\mathcal{S}(\omega))$ and the estimated distribution $\mathbb{P}(\hat{\mathcal{S}}(\omega))$. Refer to \cite{gulrajani2017improved} for the detailed justifications.
\begin{figure}[tb!]
\begin{center}
\includegraphics[width=1\linewidth]{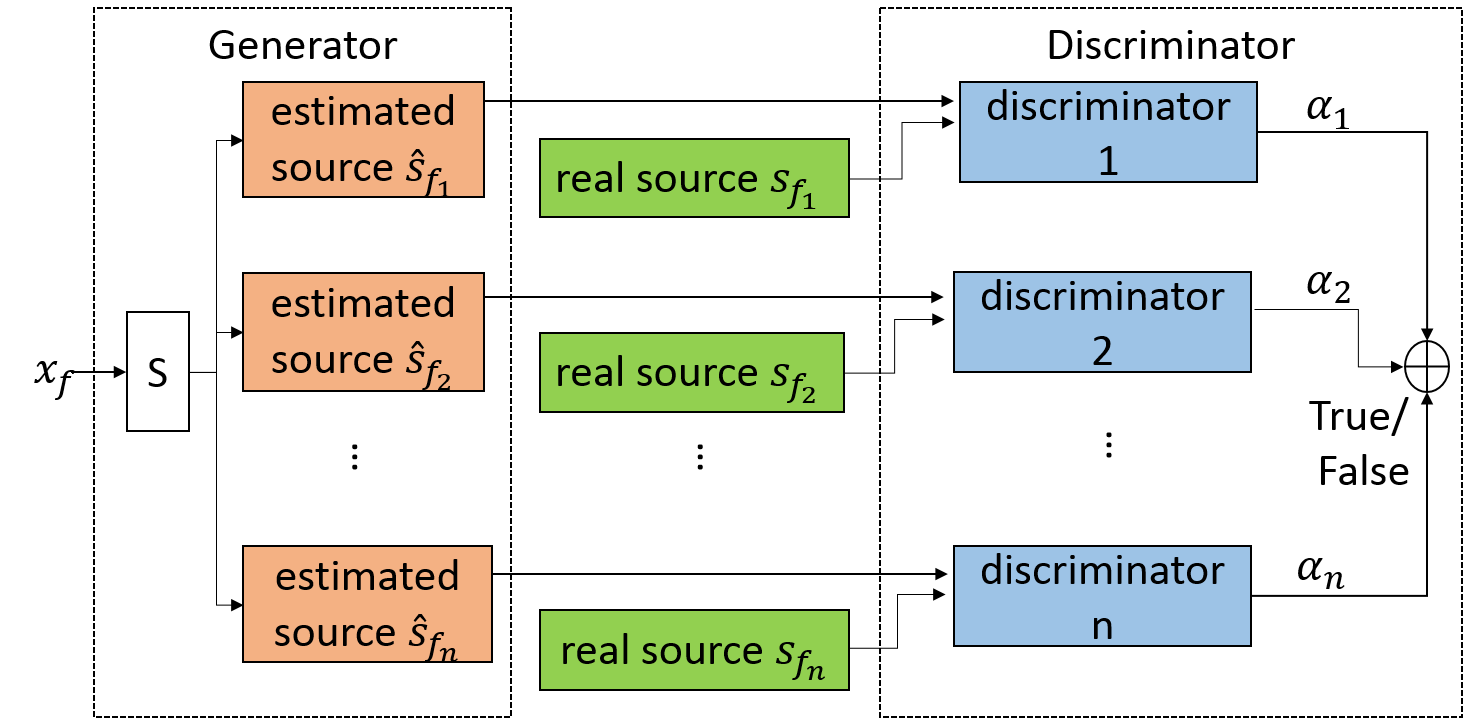}
\end{center}
\caption{Overview of the proposed framework for source separation. The input mixture signal is first transformed into the spectrum domain as $x_f$, then it is separated into $n$ sources $\hat{s}_{f_{i}}$. For each source, a discriminator is devised by using the separated i.e. generated sources as well as the real sources from auxiliary source data. The overall discriminative loss is a combination of multiple sources.}
\label{fig:overview}
\end{figure}
\subsubsection{Energy preserved Wasserstein loss}
By combing the above two loss functions, our objective function can be further written as a minmax problem whose objectives can be optimized alternatively:
\begin{small}
\begin{align}\label{eq:minmax}\notag
\min_{\theta}\max_{w}\underbrace{\mathbb{E}_{\mathcal{S}_{f} \sim \mathbb{P}(\mathcal{S}_{f})}(D_{\theta}(\mathcal{S}_{f}))-\mathbb{E}_{x_{f}(\omega) \sim \mathbb{P}(x_{f})}(D_{\theta}(G_w(x_{f}(\omega))))}_{\text{Wasserstein loss between separated and real source}} \\\notag
- \underbrace{\lambda \mathbb{E}_{\tilde{\mathcal{S}}_{f}(\omega)\sim \mathbb{P}(\tilde{\mathcal{S}}_{f}(\omega))}\left[\left(\|\nabla_{\tilde{\mathcal{S}}(\omega)}D_{\theta}(\tilde{\mathcal{S}}_{f}(\omega))\|_{2}-1\right)^2\right]}_{\text{1-Lipschitz regularizer}}\\
+ \underbrace{\beta \left(\sum_{i=1}^{n} \|\hat{s}_{f_{i}}(\omega)\|_{2}^{2} - \|x_{f}(\omega)\|_{2}^{2}\right)^{2}}_{\text{Energy preservation}}
\end{align}
\end{small}
where $\beta$ is the energy integrity penalty weight. Here we impose a Lipschitz restriction similar to~\cite{gulrajani2017improved}, which is denoted as $\|f\|_{L} \leqslant K$, for $K\geqslant 0 $ -- for all real $x_{1}$ and $x_{2}$, $|f(x_{1})-f(x_{2})|\leqslant K|x_{1}-x_{2}|$.

Note that existing audio separation models often use additional regularization terms to maximize the independence between the estimated sources. In this paper, we do not use the independence regularization explicitly. Because the independent restriction is implied in the above generative adversarial process as the estimated sources are forced to be sampled from real source distributions. This helps simplify our model.

In training stage, we randomly select mixture signals and input them to the generator $G_\theta$. The estimated sources for these mixture will then be output. These estimates, together with randomly selected real sources, are then feed into the discriminator $D_w$. Note that each kind of source was input to a specific individual discriminator $d_{w}$ for each source. We then alternately optimize $D_w$ and $G_\theta$ through gradient decent on the losses of them. After training, the generator can be taken as a music source separator.
\begin{figure}[tb!]
\begin{center}
\includegraphics[width=1\linewidth]{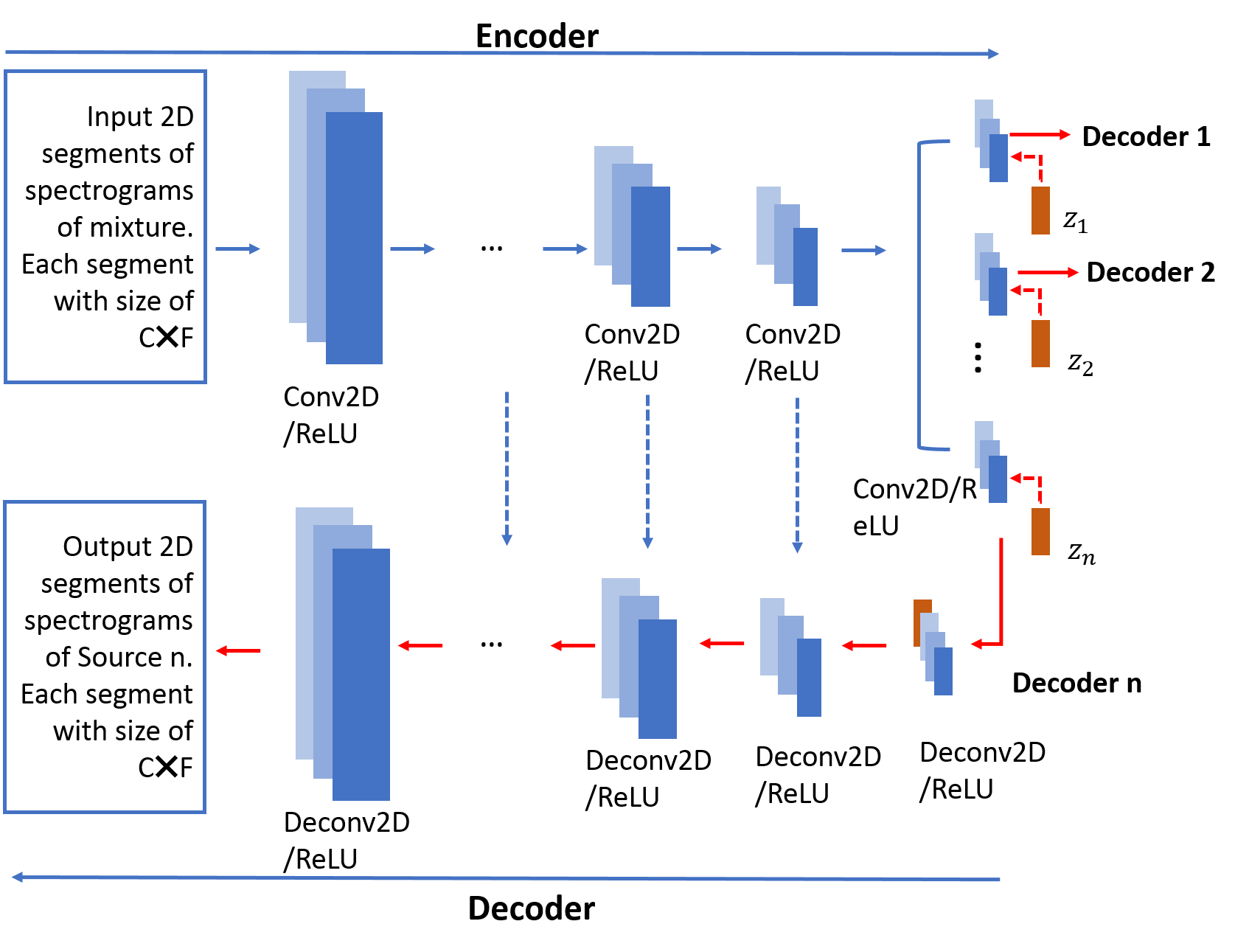}
\end{center}
\caption{Structure of the generator (only the decoder for source $n$ is depicted). The decoders for other sources share the similar structure to decoder $n$. `conv2D/ReLU' denotes a 2D convolutional layer followed by rectified linear unit (ReLU). `deconv2D/ReLU' denotes a 2D convolutional layer followed by rectified linear unit (ReLU). $C$ denotes the time context and $F$ denotes the frequency bins. $z_{i}$ is the random noises added to each output of the encoder.}
\label{fig:generator}
\end{figure}

The training procedure of the proposed generative adversarial separation model is described in Algorithm~\ref{alg:framework}.
\begin{algorithm}[tb!]
\caption{Source Separation GAN (SSGAN): Energy preserved Wasserstein learning of audio source separation.}
\label{alg:framework}
\begin{algorithmic}[1]
\REQUIRE Sectrograms of mixture audio $\{x_{f}(\cdot)\}$, and the spectrograms of source audios $s_{f_{i}(\cdot)}, i=1,...,n$
\REQUIRE $G_\theta(\cdot)$: generator; $D_w(\cdot)$: discriminator; $\beta$: energy integrity penalty weight; $\lambda$: gradient penalty coefficient; $m$: batch size; $I_{critic}$: number of discriminator iterations per generator iteration. $\omega$: time indice.
\STATE Initialize $G_\theta(\cdot)$ and $D_w(\cdot)$ with random weights.
\REPEAT
\STATE Randomly select $m$ spectrograms of mixture audios $\{x_{f}(\omega)\}, \omega=1,...,m$. Input them to the generator $G_w$. Get $n$ estimated sources for each spectrogram $\{\hat{s}_{f_{i}}(\omega)\}, \omega=1,...,m; \  i=1,...,n$.
\STATE Update generator parameters $\theta$ via optimizing Equation \ref{eq:minmax} by Adam \cite{Kingma2014Adam}.
\FOR {d-steps in $I_{critic}$}
\STATE Randomly select $m$ spectrograms of each real source $\{s_{f_{i}}(\omega)\} \sim \mathbb{P}_{s_{f_{i}}} ,\omega =1,...,m; i=1,...,n$. Randomly select $m$ spectrograms of mixture data  $\{x_{f} (\omega)\}\sim \mathbb{P}_{x}, \omega =1,...,m$ as input to $D_\theta$
\STATE  Update discriminator parameters $w$ via gradient decent by Equation \ref{eq:minmax} by Adam \cite{Kingma2014Adam}.
\ENDFOR
\UNTIL{model converges}
\end{algorithmic}
\end{algorithm}
\subsection{Architecture of ASS Generator/Discriminator}
In our model, the generator aims to separate the mixed signal into $n$ sources, while the discriminator tries to distinguish each separated source with the real ones. We use two neural networks to model both of the generator and discriminator. The working flow is illustrated in Figure~\ref{fig:overview}.

For the generator i.e. separator, we propose to use a structure of convolutional auto-encoder with one channel of input and $n$ channels for output. The network takes single channel mixture $x_{f}(\omega)$ as input, and outputs $n$ estimated signals $\{\hat{s}_{f_{i}}(\omega)\}_{i=1...n}$ with $n$ channels. Each channel corresponds to a kind of source signal. The separator is depicted in Figure~\ref{fig:generator}, which is a fully convolutional auto-encoder. In the encoding stage, the input mixture spectrograms are mapped into low dimensional features by a number of strided convolutional layers followed by rectified linear units (ReLUs). For decoding, these low dimensional features are projected back to high dimensional outputs via a number of fractional strided convolutional layers. No dense layer is used. We devise this architecture after the guidelines in~\cite{Radford2015Unsupervised}.

The encoding part of the generator aims to extract low dimensional source-specific features from the input data. Hence the encoder is devised to extract source-common features through several convolutional layers and then output $n$ source-specific features in the last layer.  Before passing these source-specific features to decoders, we add some noise to increase diversity to the output of separator. Note that the generator network consists of skip connections \cite{ronneberger2015u}, connecting the encoding layers to its corresponding decoding layers. These skip connections can pass low level details directly to higher layers, and help the decoder to reconstruct the source signals. Additionally, gradients in back propagation process can flow deeper via the skip connections.

The discriminator $D(\cdot)$ is another CNN which is larger and deeper than generator. It takes in the estimated source signals and output a similarity score between the distribution of separated sources $\mathbb{P}(\hat{s}_{f})$ and that of the real sources $\mathbb{P}(s_{f})$. For modeling flexibility, we devise separate discriminators $d_{i}(\cdot)$ for each source, and then combined their losses together to obtain the final loss: $D(\{s_{1},...,s_{n}\}) = \sum_{i}^{n}\alpha_{i}*d_{i}(s_{i})$, $d_{i}(\cdot)$ where $\alpha_{i}$ is its corresponding weight.
\section{Experiments and Discussion}
\subsection{Evaluation Protocol}
\textbf{Dataset} We term our method as source separation GAN (SSGAN). We convert the stereo songs into mono by computing the average of the two channels for all songs and sources in Demixing Secrets Dataset (DSD100)\footnote{http://liutkus.net/DSD100.zip}, which is designed to evaluate the performances of music audio source separation methods. Dataset consists of 100 professionally produced full track songs from the Mixing Secrets Free Multitrack Download Library\footnote{http://cambridge-mt.com/ms-mtk.htm}. For each song, the mixture and the sources have the same length and are all sampled at 44.1 K Hz. The DSD100 is divided into a dev set and a test set. Each of them consist of 50 songs. We train our proposed model on the dev set and test the trained model on the test set. The magnitude spectrogram for the data is computed by STFT. We use non-overlapping Hanning window~\cite{Podder2014Comparative} of 1024 points, which corresponds to about 23 milliseconds. For each frame, the first 513 FFT points were taken as magnitude values. For the input and output data for the SSGAN, we aggregate magnitude vectors in a context window of 32 frames. Hence, one SSGAN input vector has a size of $32 \times 513$, and each input and output instant spans around 743 milliseconds. All of the inputs are scaled to $[0,1]$.

This dataset also provides source tracks for \emph{drums}, \emph{vocals}, \emph{bass}, and \emph{other} instruments for each song in the set. we set generator with four output channels, and the discriminator with four input channels. Each of the channels corresponds to a kind of source. Details of the generator and discriminator networks can be found in Table~\ref{tab:generator} and Table~\ref{tab:discriminator}. The generator is composed of an encoder and four decoders. The encoder consists 5 convolutional layers followed by rectified linear units. The last layer of the encoder has four output channels, to each a random noise is added. The four decoders are of the same structure. Each step in the decoder consists of a fractional strided convolutional layers followed by rectified linear units, and a concatenation with the correspondingly encoder layer.  The four discriminators share the same structure. Each discriminator contains 6 convolutional layers followed by leaky rectified linear units, and a fully connected (FC) output layer indicating the similarity between the estimated source and the real source. The outputs from the four discriminators are weighted summed together into the final score.

\begin{table}[tb!]
\begin{center}
\resizebox{0.48\textwidth}{!}{
		\begin{tabular}{rrr}
            \toprule
parameter & description & value\\
\midrule
$C$&time context& 32\\
$F$& frequency bins & 513\\
$\alpha_{1}$& weight of discriminator of bass&0.25 \\
$\alpha_{2}$& weight of discriminator of drum&0.25 \\
$\alpha_{3}$& weight of discriminator of vocal&0.4 \\
$\alpha_{4}$& weight of discriminator of others&0.1 \\
$\beta$& energy integrity penalty weight in Eq.~\ref{eq:gans} & 10 \\
$\lambda$& gradient penalty weight in Eq.~\ref{eq:gans} & 10 \\
\bottomrule
\end{tabular}
}
\end{center}
\caption{Parameter notations and settings.}
\label{tab:paras}
\end{table}

\textbf{Model settings} The parameters for the generator and discriminator networks were initialized randomly. We used the Adam optimizer~\cite{Kingma2014Adam} with hyperparameters $\alpha=0.0001, \beta_{1}=0.5, \beta_{2}=0.9$ to train the generator and the discriminator, using a batch size of 16.

The other parameters related to the model setup can be found in Table~\ref{tab:paras}. We note that the \emph{other} source comprises many kind of instruments, and varied a lot across songs. Besides, some of the segments in the \emph{other} source is closed to the \emph{vocals} source. This indicates that the distribution of the \emph{other} source is highly dispersed, and is sometimes confounded with that of the \emph{vocals}.  To learn the distribution of the \emph{other} source well is more difficult than that of \emph{drums}, \emph{vocals},  and \emph{bass}. Hence we reduce the weight of the discriminator for the \emph{other} , and raise the weight for \emph{vocals}.

\begin{table}[tb!]
\begin{center}
\resizebox{0.48\textwidth}{!}{
\begin{tabular}{crrr}
\toprule
&layer&  $\#$ filters & output shape\\
\midrule
 \multirow{3}{*}{\rotatebox[origin=c]{90}{Encoder}}
&conv2D(3,3,2)/ReLU&16 & (16,256,16)\\
&conv2D(3,3,2)/ReLU & 32 & (8,128,32)\\
&conv2D(3,3,2)/ReLU & 64 & (4,64,64)\\
&conv2D(3,3,2)/ReLU & 128 & (2,32,128)\\
&conv2D(3,3,2) & 256*4 & (1,16,256)*4\\
&noise(1,16,1)/ReLU&---&(1,16,256)*4\\
\midrule
 \multirow{3}{*}{\rotatebox[origin=c]{90}{Decoder}}
&deconv2D(3,3,2)/ReLU & 128 & (2,32,128)/Skip1:\\
&deconv2D(3,3,2)/ReLU & 64 & (4,64,64)/Skip2:\\
&deconv2D(3,3,2)/ReLU & 64 & (8,128,32)/Skip3:\\
&deconv2D(3,3,2)/ReLU & 64 & (16,256,16)\\
&deconv2D(3,3,2)/ReLU & 1 & (32,513,1)\\
\bottomrule
\end{tabular}
}
\end{center}
\caption{The detailed structure of generator (i.e. separator). The output shape is shown as (time-context, frequency bins). `conv2D(3,3,2)' denotes 2D convolutional layer with filter size $3\times 3$, and stride = 2 in both time-context direction and frequency bin direction. `LeakyReLu(0.2)' denotes a leaky rectified linear layer with slide 0.2. 'Skip' denotes a skip connection from the corresponding encoder layer. The input data is of size 32 frames and 513 frequency bins. All the decoders share the same structure}
\label{tab:generator}

\end{table}

\begin{table}[tb!]
\begin{center}
\resizebox{0.48\textwidth}{!}{
\begin{tabular}{rrr}
\toprule
layer&  $\#$ filters & output shape\\
\midrule
conv2D(3,3,2)/ LeakyReLU(0.2) & 64 & (16,256,64)\\
conv2D(3,3,2)/ LeakyReLU(0.2) & 64 & (8,128,64)\\
conv2D(3,3,1)/ LeakyReLU(0.2)  & 128 & (8,128,128)\\
conv2D(3,3,2)/ LeakyReLU(0.2) & 128 & (4,64,128)\\
conv2D(3,3,2)/ LeakyReLU(0.2) & 256 & (2,32,256)\\
conv2D(3,3,1)/ LeakyReLU(0.2) & 256 & (2,32,256)\\
FC layer/ LeakyReLU(0.2)&---&(1)\\
\bottomrule
\end{tabular}
}
\end{center}
\caption{Detailed structure of each discriminator. The output shape is shown as (time-context, frequency bins). `conv2D(3,3,2)' denotes 2D convolutional layer with filter size $3\times 3$, and stride = 2 in both time-context direction and frequency bin direction. `LeakyReLu(0.2)' denotes a leaky rectified linear layer with slide 0.2. The input data is of size $32*513$, denoting 32 frames and 513 frequency bins.}
\label{tab:discriminator}
\end{table}

\textbf{Metrics} Following the widely used protocol, we measure the performance of the proposed separation method by the source to distortion ratio (SDR), source to interference ratio (SIR), source to artifacts ratio (SAR), source image to spatial (ISR). Among them, SDR is usually considered as the overall performance for source separation~\cite{Vincent2006Performance}.

We compute the above metrics using BSS Eval toolbox~\cite{Vincent2006Performance}. The SiSEC 2016 (Signal Separation Evaluation Campaign) published the evaluation results of all the submitted methods\footnote{https://sisec17.audiolabs-erlangen.de/\#/methods}. This facilitates the comparison of our SSGAN based method with others. We note that the published results did not cover all the 50 songs in the test set. Only the results of 47 songs are published. To make it fair, we also compute our results on the same 47 songs in the test set.

\textbf{Compared weakly supervised methods} We choose three state-of-the-art weakly supervised methods: kernel additive
modellings (KAM1 and KAM2) \cite{liutkus2014kernel,Liutkus2015Scalable}, and OZE (as termed by the SiSEC 2016 -- Signal Separation Evaluation Campaign https://sisec.inria.fr/) \cite{ozerov2012general} for comparison as our method is also weakly supervised. Note that the three peer methods are not based on deep network. In fact we have not identified any weakly supervised source separation deep network during the production of this paper, until a concurrent work in \cite{StollerICASSP18}.

\begin{figure}[tb!]
\centering
\includegraphics[width=1\linewidth]{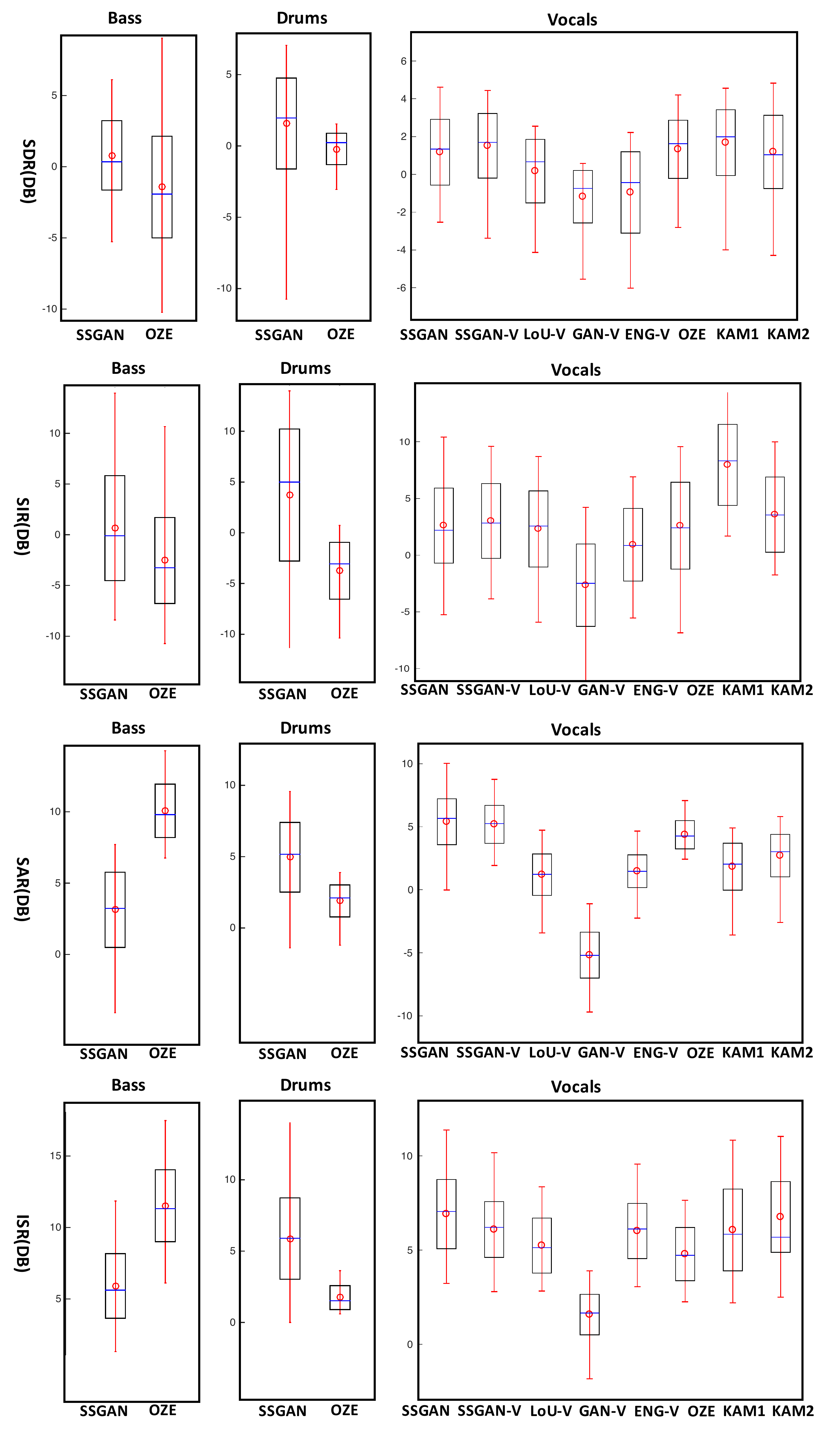}
\caption{The overall performance (SDR) and the three individual metrics (SIR, SAR, ISR) for \emph{bass}, \emph{drums}, \emph{vocals} separation task and the \emph{vocals} vs. \emph{non-vocals} separation task on the test part of DSD100. The ends of red vertical lines denote the maximum and minimum values, and the red $\circ$ denote the mean values. The upper and lower edges of the boxes denote the $mean \pm standard \ deviation$. The horizontal line in the box denotes the median values.}
\label{fig:results}
\end{figure}

\subsection{Results and Discussion}
The performance of peer methods are all from SiSEC Competition 2016 results\footnote{https://sisec17.audiolabs-erlangen.de/\#/results/1/4/2}. As the peer methods KAM1 and KAM2 \cite{Liutkus2015Scalable} only disclose performance on \emph{vocals} vs. \emph{non-vocals} separation task, hence we also test this case and denote as SSGAN-V (for \emph{vocals}) for our method. The full separation result and the binary \emph{vocals} vs. \emph{non-vocals} separation are both shown in Figure \ref{fig:results}. For our approach, the full separation and two-source separation models are separately trained by setting different number of sources and discriminators.

To prove the superiority of the proposed model. We also test the effect of the U-net structure. For the \emph{vocals} vs. \emph{non-vocals} separation task, we test the performance when the proposed model lacking of U-net structure, denoted as LoU-V. Besides, to verify the complementary effect of the energy-preserving loss and the wasserstein loss, we test the performance of the \emph{vocals} vs. \emph{non-vocals} separation task when the energy-preserving loss is removed, denoted as GAN-V, and when only the energy-preserving loss is used, denoted as ENG-V. These experimental results are shown in Figure \ref{fig:results}.

Our model performs competitively against the peer methods. In particular, SSGAN's overall performance SDR outperforms on \emph{bass} and \emph{drums}. The average SDR improvements of the proposed methods over OZE method is approximately 1.32 dB. A comparison test using a non-parametric Mann-Whitney U test, at the $5\%$ confidence interval level, shows that the proposed method is significantly better than the OZE method on the \emph{drum}and \emph{bass} channels, while it shows that the difference on the \emph{vocals} channel is not significant.
We also note our binary separation version SSGAN-V slightly improves the full version SSGAN on the \emph{vocals} vs. \emph{non-vocals} separation task in terms of the overall performance SDR. For SDR on \emph{vocals}, the performance difference is not significant. We conjecture this is because learning \emph{vocals}'s real data's distribution is more challenging to learning its Wasserstein distance as the \emph{vocals} can contain diverse patterns.

When the U-net connections are removed, the performances of LoU-V dropped obviously compared to that of SSGAN-V. This demonstrates the effect of the U-net connections. The performances of GAN-V and ENG-V are significantly lower than that of SSGAN-V. This proves our idea that the wasserstein loss and the energy-preserving loss are complementary. The energy-preserving loss can constrain the multiple sources as a whole while the wasserstein loss can constrain each source separately to make each of them look more realistic. There is quite obviously synergy between the two losses.

It is important to note that the OZE method~\cite{ozerov2012general} leverages auxiliary RWC dataset~\cite{Goto2003RWC} which contains a large amount of independent \emph{bass} and \emph{drums} source data to pretrain their model. Comparatively we only use the relatively small data from DSD100 dev (50 songs), which may cause the difference in performance. The KAM model~\cite{liutkus2014kernel} also requires deliberate model selection to fit with the source data which is no need in our method. Without such fine-tuning, the performance of KAM may drop as discussed in their paper~\cite{liutkus2014kernel}.

\section{Conclusion}
This paper is an endeavor to explore weakly supervised neural network methods for source separation, which is a relatively less studied compared with the dominant supervised learning based network models. As source separation is an ill-posed problem we resort to adversarial learning which has not been well explored for source separation in literature. We devise a loss involving spectrum energy preservation term and Wasserstein loss between the separated sources and the real source data. Our model is easy to implement and free from deliberate design of ad-hoc constraints and assumptions on the sources, thus being promising for practical applications.
\small
\bibliographystyle{named}
\bibliography{main}
\end{document}